\documentclass{plain}
\usepackage{graphicx}
\usepackage{supertabular,lscape,epsfig}
\usepackage{amssymb}
\usepackage{amsmath}
\usepackage{wasysym}
\SetPages{1}{0}

\SetVol{0}{2023}

\begin{document}

\begin{Titlepage}
\Title{Search for planets in hot Jupiter systems with multi-sector TESS photometry. III. A study of ten systems enhanced with new ground-based photometry\footnote{This research is partly based on: (1) data obtained at the 1.5 m and 0.9 m telescopes of the Sierra Nevada Observatory (Spain), which is operated by the Consejo Superior de Investigaciones Cient\'{\i}ficas (CSIC) through the Instituto de Astrof\'{\i}sica de Andaluc\'{\i}a; (2) observations made with the Liverpool Telescope operated on the island of La Palma by the Liverpool John Moores University in the Spanish Observatorio del Roque de los Muchachos of the Instituto de Astrofisica de Canarias with financial support from the UK Science and Technology Facilities Council; and (3) observations obtained with telescopes of the University Observatory Jena, which is operated by the Astrophysical Institute of the Friedrich-Schiller-University.}}
\Author{G.~~M~a~c~i~e~j~e~w~s~k~i$^{1}$, ~ M.~~F~e~r~n~\'a~n~d~e~z$^2$, ~ A.~~S~o~t~a$^2$, ~ P.~J.~~A~m~a~d~o$^2$, ~ J.~~O~h~l~e~r~t$^{3,4}$, ~ R.~~B~i~s~c~h~o~f~f$^{5}$, ~ W.~~S~t~e~n~g~l~e~i~n$^{5}$, ~ M.~~M~u~g~r~a~u~e~r$^{5}$, ~ K.-U.~~M~i~c~h~e~l$^{5}$, ~ J.~~G~o~l~o~n~k~a$^{1}$, ~ A.~~B~l~a~n~c~o~~S~o~l~s~o~n~a$^6$, ~ E.~~L~a~p~e~{\~n}~a$^6$, ~ J.~~M~o~l~i~n~s~~F~r~e~i~r~e$^6$, ~ A.~~D~e~~l~o~s~~R~\'{\i}~o~s~~C~u~r~i~e~s~e~s$^6$, ~ J.~A.~~T~e~m~p~r~a~n~o~~S~i~c~i~l~i~a$^6$}
{$^{1}$Institute of Astronomy, Faculty of Physics, Astronomy and Informatics,
         Nicolaus Copernicus University in Toru\'n, Grudziadzka 5, 87-100 Toru\'n, Poland,
         e-mail: gmac@umk.pl\\
 $^{2}$Instituto de Astrof\'isica de Andaluc\'ia (IAA-CSIC), Glorieta de la Astronom\'ia 3, 18008 Granada, Spain\\
 $^{3}$Michael Adrian Observatorium, Astronomie Stiftung Trebur, 65428 Trebur, Germany\\
 $^{4}$University of Applied Sciences, Technische Hochschule Mittelhessen, 61169 Friedberg, Germany\\
 $^{5}$Astrophysikalisches Institut und Universit\"ats-Sternwarte, Schillerg{\"a}{\ss}chen 2, 07745 Jena, Germany\\
 $^{6}$Valencia International University, 46002 Valencia, Spain}

\Received{June 2023, accepted for publication in Acta Astronomica vol. 73}
\end{Titlepage}

\Abstract{The loneliness of hot Jupiters supports the high-eccentricity migration as a primary path leading to the formation of systems with those planets stripped of any close-in planetary companions. Here we present the null results of searches for low-mass planets close to hot Jupiters in 10 planetary systems: HAT-P-4, HAT-P-10, HAT-P-12, HAT-P-17, HAT-P-19, HAT-P-32, HAT-P-44, Qatar-6, TrES-4, and WASP-48. We employed multi-sector time-series photometry from the Transiting Exoplanet Survey Satellite enhanced with new ground-based transit light curves to determine the sizes of hypothetical planets that might still avoid being detected. We redetermined transit parameters for the known hot Jupiters using a homogenous approach. We refuted transit timing variations for HAT-P-12~b, claimed recently in the literature. The transit timing data permitted us to place tighter constraints on third bodies in HAT-P-19 and HAT-P-32 systems detected in Doppler measurements. We also study four multi-periodic pulsating variable stars in the field around HAT-P-17.}{Hot Jupiters -- Stars: individual: HAT-P-4, HAT-P-10, HAT-P-12, HAT-P-17, HAT-P-19, HAT-P-32, HAT-P-44, Qatar-6, TrES-4, WASP-48, BD+30 4487, Gaia DR3 11849720750852486656, TYC 2717-453-1, Gaia DR3 1849743737517511296 -- Planets and satellites: individual: HAT-P-4~b, HAT-P-10~b, HAT-P-12~b, HAT-P-17~b, HAT-P-19~b, HAT-P-32~b, HAT-P-44~b, Qatar-6~b, TrES-4~b, WASP-48~b}


\section{Introduction}

Orbital architectures of planetary systems with massive planets on tight orbits, so-called hot Jupiters, shed unique light on the mechanisms of planetary formation. Unless those giant planets were formed in situ (Batygin \etal 2016), they must be transferred from their birthplace beyond the water-ice line into their current orbits. Whilst the early migration in a proto-planetary disk and in-situ formation can preserve nearby low-mass planets, the violent evolution paths leave hot Jupiters in seclusion (Mustill \etal 2015). Statistical studies show that the latter scenario is more probable because the hot Jupiters usually constitute single-planet systems or are accompanied by other massive planets on wide orbits. However, the observed lack of eccentric proto-hot Jupiters (Dawson \etal 2015) and a slowly growing number of hot Jupiters in compact orbital configurations (\eg Hord \etal 2022) show that the actual statistics still need clarification.

In our project (Maciejewski 2020, 2022), we use publicly available photometric data acquired with the Transiting Exoplanet Survey Satellite (TESS, Ricker \etal 2015) to search for additional planets with the transit method. In addition, we homogeneously analyse transit times for known hot Jupiters to search for perturbations induced by potential non-transiting planets close to resonant configurations (\eg Wu \etal 2023).

This paper presents results for 10 planetary systems: HAT-P-4, HAT-P-10, HAT-P-12, HAT-P-17, HAT-P-19, HAT-P-32, HAT-P-44, Qatar-6, TrES-4, and \mbox{WASP-48}. The space-borne observations were enhanced with new transit light curves acquired with ground-based instruments. For 4 systems, we acquired additional photometric time series, which complemented the transit search based on the TESS data.


\section{Systems of the sample}

The basic observational properties of the investigated systems are presented in Table~1. Their stellar and planetary properties are collected in Table~2. Below, a brief characterisation of the individual systems is provided.

\MakeTable{lcccccc}{12.5cm}{Observational properties of the systems of the sample} 
{\hline
System  & RA (J2000)  & Dec (J2000) & $m_{\rm G}$  & Distance & $d_{\rm tr}$      & $\delta_{\rm tr}$ \\
        & hh:mm:ss.s  & $\pm$dd:mm:ss   & (mag)    & (pc)     & (min)             & (ppth)            \\
\hline 
HAT-P-4  & 15:19:57.9 &  +36:13:47 & 11.1 & $321.8\pm1.6$ & $253.2^{+2.0}_{-3.2}$ &  $7.64^{+0.18}_{-0.19}$ \\
HAT-P-10 & 03:09:28.5 &  +30:40:25 & 11.6 & $125.1\pm2.1$ & $157.1^{+3.4}_{-3.7}$ & $17.16^{+0.49}_{-0.44}$ \\
HAT-P-12 & 13:57:33.5 &  +43:29:37 & 12.4 & $141.9\pm0.2$ & $140.7^{+5.4}_{-4.2}$ & $20.12^{+0.69}_{-0.73}$ \\
HAT-P-17 & 21:38:08.7 &  +30:29:19 & 10.3 & $ 92.4\pm0.2$ & $243.4^{+6.4}_{-5.5}$ & $15.06^{+0.51}_{-0.52}$ \\
HAT-P-19 & 00:38:04.0 &  +34:42:42 & 12.5 & $201.7\pm0.6$ & $166.2^{+7.0}_{-5.0}$ & $19.30^{+0.69}_{-0.84}$ \\
HAT-P-32 & 02:04:10.3 &  +46:41:16 & 11.1 & $286.2\pm1.7$ & $184.9^{+2.4}_{-2.0}$ & $23.23^{+0.27}_{-0.27}$ \\
HAT-P-44 & 14:12:34.6 &  +47:00:53 & 13.0 & $348.4\pm1.7$ & $183.7^{+5.7}_{-4.9}$ & $18.79^{+0.48}_{-0.47}$ \\
Qatar-6  & 14:48:50.5 &  +22:09:09 & 11.1 & $101.0\pm0.2$ & $95.7^{+4.4}_{-5.1}$  &  $22.9^{+1.5}_{-1.0}$   \\
TrES-4   & 17:53:13.0 &  +37:12:43 & 11.5 & $523.7\pm7.1$ & $214.7^{+9.0}_{-8.9}$ &  $9.53^{+0.27}_{-0.23}$ \\
WASP-48  & 19:24:39.0 &  +55:28:23 & 10.8 & $462.1\pm2.2$ & $186.1^{+6.4}_{-5.2}$ &  $9.06^{+0.24}_{-0.24}$ \\
\hline
\multicolumn{7}{l}{Coordinates were taken from the Gaia Data Release 3 (DR3, Gaia Collaboration \etal 2021).}  \\
\multicolumn{7}{l}{$m_{\rm G}$ is the apparent brightness in the $G$ band from the DR3. Distance is calculated on the}  \\
\multicolumn{7}{l}{DR3 parallaxes with uncertainties from error propagation. $d_{\rm tr}$ and $\delta_{\rm tr}$ are the transit duration}  \\
\multicolumn{7}{l}{and transit depth refined in this study. ppth stands for parts per thousand of the normalised}  \\
\multicolumn{7}{l}{out-of-transit flux.}  \\
}

\MakeTable{lccccc}{12.5cm}{Physical properties of the systems of the sample} 
{\hline
Star   & $M_{\star}$ $(M_{\odot})$   & $R_{\star}$ $(R_{\odot})$ & $T_{\rm eff}$ (K) & $\log g_{\star}$ ($\rm{cm \, s^{-2}}$) & [Fe/H]  \\
\hline 
HAT-P-4  & $1.26^{+0.06}_{-0.14}$ & $1.59\pm0.07$             & $5860\pm80$  & $4.14^{+0.01}_{-0.04}$ & $+0.24\pm0.08$   \\
HAT-P-10 & $0.83 \pm 0.03$        & $0.79\pm0.02$             & $4980\pm60$  & $4.56\pm0.02$          & $+0.13\pm0.08$   \\
HAT-P-12 & $0.733 \pm 0.018$      & $0.701^{+0.017}_{-0.012}$ & $4650\pm60$  & $4.61\pm0.01$          & $-0.29\pm0.05$   \\
HAT-P-17 & $0.857 \pm 0.039$      & $0.838 \pm 0.21$          & $5246\pm80$  & $4.52\pm0.02$          & $ 0.00\pm0.08$   \\
HAT-P-19 & $0.842 \pm 0.042$      & $0.820 \pm 0.048$         & $4990\pm130$ & $4.54\pm0.05$          & $+0.23\pm0.08$   \\
HAT-P-32 & $1.160 \pm 0.041$      & $1.219 \pm 0.016$         & $6207\pm88$  & $4.33\pm0.01$          & $-0.04\pm0.08$   \\
HAT-P-44 & $0.942 \pm 0.041$      & $0.949^{+0.080}_{-0.037}$ & $5295\pm100$ & $4.46\pm0.06$          & $+0.33\pm0.10$   \\
Qatar-6  & $0.822 \pm 0.021$      & $0.822 \pm 0.021$         & $5052\pm66$  & $4.64\pm0.01$          & $-0.025\pm0.094$ \\
TrES-4   & $1.22 \pm 0.17$        & $1.738 \pm 0.092$         & $6100\pm100$ & $4.045\pm0.034$        & $0.0\pm0.2$      \\
WASP-48  & $1.09 \pm 0.08$        & $1.09 \pm 0.14$           & $6000\pm150$ & $4.50\pm0.15$          & $-0.12\pm0.12$   \\
\hline 
Planet     & $M_{\rm b}$ $(M_{\rm Jup})$ & $R_{\rm b}$ $(R_{\rm Jup})$ & \multicolumn{3}{l}{Data source} \\
\hline 
HAT-P-4 b  & $0.68 \pm 0.04$   & $1.27\pm0.05$             & \multicolumn{3}{l}{Kov\'acs \etal (2007)} \\
HAT-P-10 b & $0.487 \pm 0.018$ & $1.005^{+0.032}_{-0.027}$ & \multicolumn{3}{l}{Bakos \etal (2009)} \\
HAT-P-12 b & $0.211 \pm 0.012$ & $0.959^{+0.029}_{-0.021}$ & \multicolumn{3}{l}{Hartman \etal (2009)} \\
HAT-P-17 b & $0.534 \pm 0.018$ & $1.010 \pm 0.029$         & \multicolumn{3}{l}{Howard \etal (2012)} \\
HAT-P-19 b & $0.292 \pm 0.018$ & $1.132 \pm 0.072$         & \multicolumn{3}{l}{Hartman \etal (2011a)} \\
HAT-P-32 b & $0.86 \pm 0.16$   & $1.789 \pm 0.025$         & \multicolumn{3}{l}{Hartman \etal (2011b)} \\
HAT-P-44 b & $0.352 \pm 0.029$ & $1.242^{+0.106}_{-0.051}$ & \multicolumn{3}{l}{Hartman \etal (2014)} \\
Qatar-6 b  & $0.668 \pm 0.066$ & $1.062 \pm 0.071$         & \multicolumn{3}{l}{Alsubai \etal (2018)} \\
TrES-4 b   & $0.84 \pm 0.10$   & $1.674 \pm 0.094$         & \multicolumn{3}{l}{Mandushev \etal (2007)} \\
WASP-48 b  & $0.98 \pm 0.09$   & $1.67 \pm 0.10$           & \multicolumn{3}{l}{Enoch \etal (2011)} \\
\hline
\multicolumn{6}{l}{$M_{\star}$, $R_{\star}$, $T_{\rm eff}$, $g_{\star}$, and [Fe/H] are the mass, radius, effective temperature, gravitational acceleration }  \\
\multicolumn{6}{l}{in cgs, and metallicity for the host star, respectively. $M_{\rm b}$ and $R_{\rm b}$ are the mass and radius of the }  \\
\multicolumn{6}{l}{transiting planet in Jupiter units. The data source refers to the parameters of both the star and } \\
\multicolumn{6}{l}{the planet.} \\
}

\textit{HAT-P-4}. This planetary system comprises a low-density, inflated planet and a metal-rich, evolved main-sequence late F or early G star (Kov\'acs \etal 2007). Winn \etal (2011) showed that the planet's 3.05-day orbit is prograde with a sky-projected angle between the planetary orbital and stellar axes of \mbox{$\lambda = -5^{\circ} \pm 12^{\circ}$}. Furthermore, a linear trend in radial velocities (RVs) was detected and identified as a constant acceleration of the systemic barycentre $\dot{\gamma} = 0.0246 \pm 0.0026 \, {\rm m \, s^{-1} \, day^{-1}}$ due to a third body (a planet or a companion star). The system parameters were refined by Christiansen \etal (2011), who analysed photometric time series acquired with the NASA EPOXI Mission of Opportunity. Then, the transit light curves were reanalysed by Southworth (2011). The system parameters were redetermined again by Wang \etal (2021). Todorov \etal (2013) obtained occultation light curves with the warm \textit{Spitzer Space Telescope} and found that heat recirculation from the day to the night side of the planet is inefficient. Furthermore, occultation timing placed a tight constraint on zero orbital eccentricity. This finding was confirmed in the RV study by Bonomo \etal (2017). Mugrauer \etal (2014) used common proper motions and radial velocities to find that the host star is likely accompanied by a widely separated G2 star, named HAT-P-4 B. Saffe \etal (2017) found the companion has metallicity lower by $\sim$0.1 dex compared to that of HAT-P-4~A. They postulated that the giant planet's migration could trigger the fall of planetesimals and rocky planets (with a total mass of $\sim$10 $M_{\oplus}$) onto HAT-P-4~A at the time of the system's formation, enriching the star with metals.
   
\textit{HAT-P-10}. The planet in this system is another low-density hot Jupiter announced by the HATNet survey (Bakos \etal 2009) and then independently by the WASP project (West \etal 2009). Thus, it is also known as WASP-11~b. It orbits a K dwarf every 3.7 d. Different approaches of both teams resulted in divergent values of the mean planetary density of $0.594 \pm 0.052$ ${\rm g \, cm^{-3}}$ (Bakos \etal 2009) and $0.926^{+0.09}_{-0.15}$ ${\rm g \, cm^{-3}}$ (West \etal 2009). In their follow-up study, Wang \etal (2014) obtained the value of $0.697^{+0.046}_{-0.062}$ ${\rm g \, cm^{-3}}$. Then, Mancini \etal (2015) found the density of $0.672 \pm 0.037$ ${\rm g \, cm^{-3}}$. Those authors also determined a spin-orbit alignment with $\lambda$ equal to $7^{\circ} \pm 5^{\circ}$. Knutson \etal (2014) postulated that the system accelerates with the rate of $-0.014 \pm 0.032$ ${\rm m \, s^{-1} \, day^{-1}}$. Ngo \etal (2015) demonstrated that a common proper motion M dwarf companion could induce this RV trend. Mugrauer (2019) showed that the system is a member of a hierarchical triple-star system.

\textit{HAT-P-12}. The planet was found to transit a K4 dwarf every 3.2 days (Hartman \etal 2009). This is a low-density ($\approx$$0.3$ ${\rm g \, cm^{-3}}$) globe with a sub-Saturn mass. The sky-projected orbital obliquity angle remains not well constrained with $\lambda = -54^{+41}_{-13}$ degrees (Mancini \etal 2018). The system's parameters were refined in the follow-ups studies by Lee \etal (2012), Mallonn \etal (2015b), Sada \& Ram\'on-Fox (2016), Mancini \etal (2018), \"{O}zt\"{u}rk \& Erdem (2019), and Wang \etal (2021). Although Sing \etal (2016) found a strong optical scattering slope from blue to near-IR wavelengths, Line \etal (2013), Mallonn \etal (2015b), Turner \etal (2017), and Yan \etal (2020) concluded that HAT-P-12~b is covered by a cloudy atmosphere ruling out the presence of the Rayleigh scattering. The scenario invoking a completely clear atmosphere was also refuted by Alexoudi \etal (2018). However, those authors found the Rayleigh scattering slope discernible in the visible transmission spectrum. Finally, Wong \etal (2020) detected both the clouds inferred from weakened water vapour absorption and Rayleigh scattering produced by small particles. Atmospheric models with photochemical hazes composed of soot or tholins were found to reproduce the planetary transmission spectrum. However, Jiang \etal (2021) noticed that conflicting results of previous atmospheric studies could be rendered by stellar contamination of unocculted stellar spots and faculae. Wong \etal (2020) also provided evidence that heat in the planet's atmosphere is efficiently redistributed between the day and night hemispheres.

\textit{HAT-P-17}. The host star is an early K dwarf, which is orbited by a half-Jupiter-mass transiting planet, HAT-P-17~b, on a 10.3-day eccentric orbit (Howard \etal 2012) and a $\approx$$3 \, M_{\rm Jup}$ planet, HAT-P-17~c, on a $\approx$4000 day orbit (Howard \etal 2012, Bonomo \etal 2017). Initially, the orbit of planet b was found to be aligned (Fulton \etal 2013), but then a slight misalignment with $\lambda = -27.5^{\circ} \pm 6.7^{\circ}$ was derived (Mancini \etal 2022). The system's parameters were refined by Mancini \etal (2022), who enhanced their analysis with TESS photometric time series acquired for two transits in sector 15.

\textit{HAT-P-19}. The K1 dwarf hosts a low-density Saturn-mass planet (Hartman \etal 2011a), moving on a 4.0-day circular orbit (Bonomo \etal 2017). The systemic barycentre accelerates outward (Hartman \etal 2011a) with a rate of \mbox{$0.440 \pm 0.061$ ${\rm m \, s^{-1} \, day^{-1}}$}, revealing the presence of a third body. The system parameters were refined by Mallonn \etal (2015a), Seeliger \etal (2015), Ba{\c{s}}t{\"u}rk \etal (2020), and Wang \etal (2021). The analysis of transmission spectra by Mallonn \etal (2015a) opts for a flat-spectrum (grey) atmosphere of HAT-P-19~b.

\textit{HAT-P-32}. The transiting planet is a low-density hot Jupiter whipping around its highly active late-F main-sequence star in 2.15 days (Hartman \etal 2011b). Its orbit is circular (Bonomo \etal 2017) and polar with the sky-projected orbital obliquity angle of $85.0^{\circ} \pm 1.5^{\circ}$ (Albrecht \etal 2012). The systemic properties were revised in studies by Seeliger \etal (2014), Tregloan-Reed \etal (2018), Wang \etal (2019), and Ba{\c{s}}t{\"u}rk \etal (2022). The presence of a third body was inferred from the linear RV trend of $-0.094 \pm 0.023$ ${\rm m \, s^{-1} \, day^{-1}}$ (Knutson \etal 2014, Bonomo \etal 2017). Furthermore, Adams \etal (2013) detected an M1.5-dwarf companion (Zhao \etal 2014), which was confirmed to be gravitationally bound by Ngo \etal (2015). However, that stellar companion was too distant to explain the observed RV trend. The analyses of transmission spectra and multicolour broad-band transit light curves show that the planetary spectrum is flat, which could come from grey absorption in clouds of the upper planetary atmosphere (Gibson \etal 2013, Nortmann \etal 2016, Mallonn \etal 2016). Mallonn \& Strassmeier (2016) and Alam \etal (2020) confirmed the clouds or hazes. Damiano \etal (2017) reported the presence of water vapour,  then confirmed by Alam \etal (2020). Mallonn \etal (2019b) placed an upper constraint on the planetary geometric albedo, which must be lower than 0.2. Czesla \etal (2022) provided hints that the planet is losing its mass via the first Lagrange point due to the Roche lobe overflow.

\textit{HAT-P-44}. Transits of HAT-P-44~b are observed every 4.3 days (Hartman \etal 2014). The planet is a bloated hot Saturn with a non-transiting massive planetary companion on a wide orbit. The orbital eccentricity of planet b and the orbital parameters of planet c remain poorly constrained in the current RV data. Mallonn \etal (2019a) acquired a new transit light curve and combined it with amateur data to refine the transit ephemeris for HAT-P-44~b. Then, Ivshina \& Winn (2022) extracted transit times from TESS sectors 16 and 23 to make this ephemeris even more precise.

\textit{Qatar-6}. The system comprises a hot sub-Jovian-mass planet on a circular 3.5-day orbit and an early-K dwarf (Alsubai \etal 2018). The orbital geometry results in grazing transits. Rice \etal (2023) showed that the orbit is well aligned. Mugrauer (2019) detected a candidate stellar companion bound to the host star. Rice \etal (2023) demonstrated that both stars likely constitute an edge-on binary system, which implies a configuration of spin-orbit and orbit-orbit alignment.

\textit{TrES-4}. TrES-4~b is a low-density, bloated hot Jupiter moving on a 3.55-day circular and aligned orbit around an F-type main-sequence star (Mandushev \etal 2007, Sozzetti \etal 2009, Bonomo \etal 2017, Narita \etal 2010). The system properties were refined in follow-up studies by Chan \etal (2011) and Sozzetti \etal (2015).

\textit{WASP-48}. The slightly evolved late-F dwarf hosts a bloated Jupiter-like planet moving along a 2.1-day circular orbit (Enoch \etal 2011, Bonomo \etal 2017). The system's parameters were revised by Ciceri \etal (2015). In their studies, O'Rourke \etal (2014), Clark \etal (2018), and Murgas \etal (2017) detected the planet's thermal emission in the infrared. They interpreted a flat optical transmission spectrum as a manifestation of a cloud-free atmosphere with titanium oxide and vanadium oxide.


\section{Observations and data reduction}

\subsection{TESS photometric time series}

\MakeTable{cccccccc}{12.5cm}{Details on TESS observations used} 
{\hline
Sect./ & from--to & $pnr$ & $N_{\rm tr}$ & Sect./ & from--to & $pnr$ & $N_{\rm tr}$ \\
/Mode  &          & (ppth) &              & /Mode  &          & (ppth) &              \\
\hline 
\multicolumn{4}{c}{HAT-P-4}                 & \multicolumn{4}{c}{HAT-P-44}                \\
24/SC & 2020-Apr-16--2020-May-13 & 2.53 & 9 & 16/LC & 2019-Sep-11--2019-Oct-07 & 5.35 & --\\
50/SC & 2022-Mar-26--2022-Apr-22 & 2.42 & 5 & 23/SC & 2020-Mar-18--2020-Apr-16 & 5.85 & 6\\
51/SC & 2022-Apr-22--2022-May-18 & 2.70 & 4 & 49/SC & 2022-Feb-26--2022-Mar-26 & 6.36 & 6\\
\multicolumn{3}{r}{total:}             & 18 & 50/SC & 2022-Mar-26--2022-Apr-22 & 6.12 & 5\\
\multicolumn{4}{c}{HAT-P-10}                & \multicolumn{3}{r}{total:}             & 17\\
18/LC & 2019-Nov-02--2019-Nov-27 & 2.85 & -- & \multicolumn{4}{c}{Qatar-6}               \\
42/SC & 2021-Aug-20--2021-Sep-16 & 2.82 & 6 & 50/SC & 2022-Mar-26--2022-Apr-22 & 2.63 & 5\\
58/SC & 2022-Oct-29--2022-Nov-26 & 2.45 & 7 & 51/SC & 2022-Apr-22--2022-May-18 & 3.15 & 4\\
\multicolumn{3}{r}{total:}             & 13 & \multicolumn{3}{r}{total:}              & 9\\
\multicolumn{4}{c}{HAT-P-12}            & \multicolumn{4}{c}{TrES-4}                     \\
16/LC & 2019-Sep-11--2019-Oct-07 & 4.08 & -- & 25/SC & 2020-May-13--2020-Jun-08 & 2.87 & 6\\
23/SC & 2020-Mar-18--2020-Apr-16 & 3.99 & 5 & 26/SC & 2020-Jun-08--2020-Jul-04 & 2.96 & 6\\
49/SC & 2022-Feb-26--2022-Mar-26 & 3.61 & 7 & 40/SC & 2021-Jun-24--2021-Jul-23 & 2.91 & 8\\
50/SC & 2022-Mar-26--2022-Apr-22 & 3.88 & 4 & 52/SC & 2022-May-18--2022-Jun-13 & 2.54 & 6\\
\multicolumn{3}{r}{total:}             & 16 & 53/SC & 2022-Jun-13--2022-Jul-09 & 2.91 & 6\\
\multicolumn{4}{c}{HAT-P-17}                & \multicolumn{3}{r}{total:}             & 32\\
15/SC & 2019-Aug-15--2019-Sep-11 & 1.55 & 2 & \multicolumn{4}{c}{WASP-48}                \\
55/SC & 2022-Aug-05--2022-Sep-01 & 1.35 & 2 & 14/LC & 2019-Jul-18--2019-Aug-15 & 3.14 & --\\
56/SC & 2022-Sep-01--2022-Sep-30 & 1.27 & 3 & 15/LC & 2019-Aug-15--2019-Sep-11 & 2.95 & --\\
\multicolumn{3}{r}{total:}              & 7 & 16/LC & 2019-Sep-11--2019-Oct-07 & 2.63 & --\\
\multicolumn{4}{c}{HAT-P-19}                & 23/SC & 2020-Mar-18--2020-Apr-16 & 3.23 & 10\\
17/LC & 2019-Oct-07--2019-Nov-02 & 4.83 & -- & 26/SC & 2020-Jun-08--2020-Jul-04 & 2.87 & 12\\
57/SC & 2022-Sep-30--2022-Oct-29 & 4.12 & 7 & 40/SC & 2021-Jun-24--2021-Jul-23 & 2.91 & 12\\
\multicolumn{3}{r}{total:}              & 7 & 41/SC & 2021-Jul-23--2021-Aug-20 & 2.64 & 12\\
\multicolumn{4}{c}{HAT-P-32}                & 53/SC & 2022-Jun-13--2022-Jul-09 & 2.87 & 10\\
18/LC & 2019-Nov-02--2019-Nov-27 & 2.78 & -- & 54/SC & 2022-Jul-09--2022-Aug-05 & 2.77 & 11\\
58/SC & 2022-Oct-29--2022-Nov-26 & 2.09 & 13 & 55/SC & 2022-Aug-05--2022-Sep-01 & 2.90 & 12\\
\multicolumn{3}{r}{total:}             & 13 & 56/SC & 2022-Sep-01--2022-Sep-30 & 2.45 & 12\\
      &                          &      &    & \multicolumn{3}{r}{total:}             & 91\\
\hline
\multicolumn{8}{l}{Mode specifies long cadence (LC) or short cadence (SC) photometry.}  \\
\multicolumn{8}{l}{$pnr$ is the photometric noise rate in parts per thousand (ppth) of the normalised flux per minute of}     \\
\multicolumn{8}{l}{exposure, see Fulton \etal (2011). $N_{\rm tr}$ is the number of transits qualified for this study.}     \\
}

Light curves were extracted from TESS science frames following the procedure described in detail in Maciejewski (2020). Here we give a short outline: The main portion of data was acquired with the short cadence (SC) of the 2-minute exposure time. The target pixel files were downloaded from the exo.MAST portal\footnote{https://exo.mast.stsci.edu}. In some sectors, only observations with the 30-minute exposure time were available. Those long-cadence (LC) data were extracted from full-frame images with the TESSCut\footnote{https://mast.stsci.edu/tesscut/} tool (Brasseur \etal 2019). The final light curves were obtained with the Lightkurve v1.9 package (Lightkurve Collaboration 2018). They were de-trended and normalised to unity outside the transits with the built-in Savitzky-Golay filter. Then, the light curves were visually inspected to remove time-correlated flux ramps and measurements affected by scattered light. 

The TESS observations used in this study are summarised for individual targets in Table~3.

\subsection{Ground-based transit observations}

\MakeTable{c l l l c c c c}{12.5cm}{Summary for the ground-based transit observations}
{\hline
ID & Date UT start-end & Telesc., filter  &  Airmass change  & $N_{\rm{obs}}$ & $t_{\rm{exp}}$ & $\Gamma$ & $pnr$\\
   &                   &                  &                  &                & (s)          &          & (ppth)\\
 \hline
\multicolumn{8}{c}{HAT-P-4 b}\\ 
 1 & 2015-05-06 21:59-03:15 & OSN 1.5, $R_{\rm{C}}$ & $1.17 - 1.00 - 1.18$ &  539 & 30 & 1.71 & 0.65 \\
 2 & 2016-03-28 22:35-04:42 & OSN 1.5, $R_{\rm{C}}$ & $1.66 - 1.00 - 1.06$ &  483 & 40 & 1.32 & 0.63 \\
 3 & 2018-04-19 19:25-02:25 & PIW 0.6, clear        & $1.61 - 1.05 - 1.14$ & 1251 & 17 & 3.00 & 1.08 \\
 4 & 2018-04-25 22:55-05:22 &  LT 2.0, $r'$         & $1.35 - 1.01 - 1.30$ &  315 & 40 & 1.02 & 0.61 \\
 5 & 2018-06-13 21:25-03:24 &  LT 2.0, $r'$         & $1.08 - 1.01 - 1.72$ &  295 & 40 & 1.02 & 0.67 \\
 6 & 2019-03-15 23:17-04:45 & OSN 1.5, $R_{\rm{C}}$ & $1.76 - 1.00 - 1.01$ &  768 & 20 & 2.35 & 0.76 \\
\multicolumn{8}{c}{HAT-P-10 b}\\ 
 1 & 2017-08-28 22:17-02:27 & PIW 0.6, clear        & $1.84 - 1.10$        &  456 & 25 & 2.00 & 1.40 \\
 2 & 2018-08-25 00:15-04:40 & OSN 1.5, $R_{\rm{C}}$ & $2.00 - 1.01$        &  335 & 40 & 1.32 & 0.98 \\
 3 & 2020-11-20 23:03-03:30 & OSN 0.9, $R_{\rm{C}}$ & $1.01 - 1.58$        &  284 & 50 & 1.07 & 1.46 \\
\multicolumn{8}{c}{HAT-P-12 b}\\ 
 1 & 2019-04-12 20:07-02:30 & PIW 0.6, clear        & $1.19 - 1.01 - 1.18$ &  233 & 57 & 1.00 & 1.72 \\
 2 & 2019-04-25 20:00-01:53 & PIW 0.6, clear        & $1.16 - 1.01 - 1.21$ &  348 & 57 & 1.00 & 1.74 \\
 3 & 2020-03-24 23:10-03:09 & PIW 0.6, clear        & $1.04 - 1.01 - 1.13$ &  358 & 37 & 1.50 & 1.54 \\
 4 & 2020-04-06 19:07-00:10 & PIW 0.6, clear        & $1.41 - 1.01 - 1.02$ &  453 & 37 & 1.50 & 2.12 \\
\multicolumn{8}{c}{HAT-P-17 b}\\ 
 1 & 2017-09-19 19:50-01:48 & OSN 0.9, clear        & $1.12 - 1.01 - 1.49$ &  906 & 20 & 2.61 & 0.70 \\
 2 & 2019-08-23 20:40-01:55 & JENA 0.9, clear       & $1.16 - 1.07 - 1.31$ &  324 & 45 & 1.05 & 1.17 \\
\multicolumn{8}{c}{HAT-P-19 b}\\ 
 1 & 2020-10-21 19:56-23:56 & JENA 0.9, clear       & $1.11 - 1.04 - 1.13$ &  237 & 45 & 1.07 & 1.46 \\
 2 & 2020-11-10 20:32-01:49 & OSN 1.5, clear        & $1.02 - 1.00 - 1.60$ &  690 & 20 & 2.35 & 0.64 \\
\multicolumn{8}{c}{HAT-P-32 b}\\ 
 1 & 2016-12-21 17:26-23:30 & PIW 0.6, clear        & $1.03 - 1.01 - 1.39$ &  707 & 27 & 2.00 & 1.33 \\
 2 & 2017-09-20 18:31-02:15 & PIW 0.6, clear        & $1.83 - 1.01 - 1.04$ & 1379 & 27 & 2.00 & 1.97 \\
 3 & 2017-12-13 15:58-20:27 & PIW 0.6, clear        & $1.19 - 1.01 - 1.02$ &  536 & 27 & 2.00 & 1.09 \\
 4 & 2018-10-12 18:29-01:25 & PIW 0.6, clear        & $1.44 - 1.01 - 1.07$ &  830 & 27 & 2.00 & 1.38 \\
 5 & 2018-12-09 19:45-01:49 & OSN 0.9, clear        & $1.05 - 1.02 - 1.62$ &  834 & 20 & 2.31 & 0.69 \\
 6 & 2019-01-06 19:41-00:45 & OSN 0.9, clear        & $1.02 - 1.97$        & 1004 & 12 & 3.34 & 0.79 \\
 7 & 2019-10-06 20:25-01:55 & OSN 0.9, $R_{\rm{C}}$ & $1.67 - 1.02$        &  330 & 50 & 1.07 & 1.65 \\
 8 & 2020-11-09 18:28-23:19 & TRE 1.2, clear        & $1.25 - 1.00 - 1.02$ &  440 & 30 & 1.54 & 0.69 \\
\multicolumn{8}{c}{HAT-P-44 b}\\ 
 1 & 2018-02-24 20:14-04:25 & PIW 0.6, clear        & $1.86 - 1.01 - 1.05$ &  482 & 57 & 1.00 & 2.01 \\
 2 & 2018-04-08 20:18-01:27 & PIW 0.6, clear        & $1.20 - 1.01 - 1.04$ &  213 & 57 & 1.00 & 1.97 \\
 3 & 2019-02-16 20:30-04:41 & PIW 0.6, clear        & $1.96 - 1.01 - 1.04$ &  736 & 37 & 1.50 & 2.49 \\
 4 & 2019-03-31 20:40-01:02 & JENA 0.9, clear       & $1.31 - 1.00$        &  261 & 45 & 1.05 & 1.85 \\
 5 & 2019-04-30 21:36-02:59 & TRE 1.2, clear        & $1.03 - 1.00 - 1.27$ &  152 &120 & 0.47 & 1.65 \\
 6 & 2020-02-26 00:22-05:51 & OSN 1.5, clear        & $1.34 - 1.02 - 1.08$ &  708 & 25 & 2.20 & 1.01 \\
\hline
\multicolumn{8}{l}{Date yyyy-mm-dd is given for the beginning of the observing run in UT. $N_{\rm{obs}}$ is the number of}\\ 
\multicolumn{8}{l}{useful scientific exposures acquired. $t_{\rm{exp}}$ is the exposure time used. $\Gamma$ is the median number of }\\ 
\multicolumn{8}{l}{exposures per minute. $pnr$ is the photometric scatter in ppth of the normalised flux per minute}\\ 
\multicolumn{8}{l}{of observation.}\\ 
}

\addtocounter{table}{-1}

\MakeTable{c l l l c c c c}{12.5cm}{Concluded}
{\hline
ID & Date UT start-end & Telesc., filter  &  Airmass change  & $N_{\rm{obs}}$ & $t_{\rm{exp}}$ & $\Gamma$ & $pnr$\\
   &                   &                  &                  &                & (s)          &          & (ppth)\\
 \hline
\multicolumn{8}{c}{Qatar-6 b}\\ 
 1 & 2019-04-09 20:22-02:15 & PIW 0.6, clear        & $1.73 - 1.17 - 1.26$ &  696 & 27 & 2.00 & 1.39 \\
 2 & 2019-04-23 19:34-00:00 & PIW 0.6, clear        & $1.69 - 1.17$        &  379 & 37 & 1.50 & 1.51 \\
 3 & 2019-04-30 22:46-01:10 & OSN 1.5, clear        & $1.12 - 1.04 - 1.05$ &  340 & 20 & 2.35 & 0.54 \\
 4 & 2019-04-30 22:16-00:59 & JENA 0.9, clear       & $1.18 - 1.14 - 1.20$ &  150 & 45 & 1.05 & 1.00 \\
 5 & 2019-05-07 22:53-01:02 & JENA 0.9, clear       & $1.14 - 1.25$        &  126 & 45 & 1.05 & 1.26 \\
 6 & 2019-05-21 23:12-02:31 & OSN 1.5, clear        & $1.04 - 1.45$        &  468 & 20 & 2.35 & 0.69 \\
 7 & 2019-05-28 22:12-02:55 & OSN 1.5, clear        & $1.04 - 1.83$        &  796 & 15 & 2.93 & 0.96 \\
 8 & 2021-02-18 02:11-05:31 & OSN 1.5, $R_{\rm{C}}$ & $1.33 - 1.04$        &  533 & 20 & 2.69 & 0.64 \\
 9 & 2021-03-04 02:43-06:09 & OSN 1.5, $I_{\rm{C}}$ & $1.10 - 1.04 - 1.14$ &  347 & 30 & 1.69 & 0.73 \\
\multicolumn{8}{c}{TrES-4 b}\\ 
 1 & 2019-05-12 20:50-01:07 & JENA 0.9, clear       & $1.64 - 1.04$        &  253 & 45 & 1.07 & 1.39 \\
\multicolumn{8}{c}{WASP-48 b}\\ 
 1 & 2018-04-06 23:34-04:00 & TRE 1.2, clear        & $1.65 - 1.05$        &  166 & 80 & 0.68 & 0.78 \\
 2 & 2018-05-04 20:28-01:28 & PIW 0.6, clear        & $1.71 - 1.04$        &  593 & 27 & 2.00 & 1.61 \\
 3 & 2018-07-18 20:54-02:01 & TRE 1.2, clear        & $1.07 - 1.00 - 1.12$ &  205 & 80 & 0.68 & 0.99 \\
 4 & 2018-08-02 20:39-02:26 & TRE 1.2, clear        & $1.03 - 1.00 - 1.28$ &  224 & 80 & 0.68 & 1.08 \\
 5 & 2018-08-02 20:33-03:01 & OSN 0.9, clear        & $1.15 - 1.05 - 1.41$ &  692 & 30 & 1.82 & 0.83 \\
 6 & 2019-07-26 21:27-02:20 & OSN 0.9, clear        & $1.12 - 1.05 - 1.22$ &  484 & 30 & 1.67 & 0.86 \\
 7 & 2019-10-22 18:23-22:34 & JENA 0.9, clear       & $1.04 - 1.56$        &  217 & 45 & 1.05 & 1.54 \\
\hline
}

The photometric time series for exoplanetary transits were acquired between 2015 and 2020. Six instruments were engaged: 
\begin{itemize}
 \item the 2.0 m Liverpool Telescope (Steele \etal 2004) at Observatorio del Roque de los Muchachos (La Palma, Spain) and the IO:I
 near-infrared camera (Barnsley \etal 2012) -- LT 2.0,
 \item the 1.5 m Ritchey-Chr\'etien Telescope at the Sierra Nevada Observatory (OSN, Spain) equipped with a Roper Scientific VersArray 2048B CCD camera -- OSN 1.5,
  \item the 1.2 m Cassegrain telescope at the Michael Adrian Observatory (Trebur, Germany), equipped with an SBIG STL-6303E CCD camera -- TRE 1.2,
  \item the 0.9 m Ritchey-Chr\'etien Telescope at the OSN, equipped with a Roper Scientific VersArray 2048B CCD camera -- OSN 0.9,
  \item the 0.9 m telescope at the University Observatory Jena (Germany) and the Schmidt Teleskop Kamera (Mugrauer \& Berthold 2010) -- JENA 0.9,  
  \item the 0.6 m Cassegrain telescope at the Institute of Astronomy of the Nicolaus Copernicus University (Piwnice near Toru\'n, Poland), equipped with an SBIG STL-1001M (by June 2018) or FLI ML16803 (from August 2018) CCD camera -- PIW 0.6.
\end{itemize}

The telescopes were automatically or manually guided to minimise field drifts during observing runs with a precision of a few arc seconds. The instrumental set-ups were defocused, spreading starlight over many CCD pixels to reduce flat-fielding errors and minimise the dead time lost for the CCD readout. The observations were primarily performed without any filter to maximise the signal-to-noise ratio for precise transit modelling, and only occasionally were the observations acquired through a red filter. 

The observing runs were scheduled in such a way as to acquire complete light curves with the full coverage of the critical phases, such as the transit's ingress and egress. In total, 48 light curves were obtained. The details on the individual runs are given in Table~4.

Data reduction was performed with AstroImageJ software (Collins \etal 2017). The science frames were calibrated following a standard procedure involving de-biasing or dark-current correction and flat-fielding with sky-flat-field frames. Fluxes were obtained with the aperture photometry method with the aperture radius and ensemble of comparison stars optimised in trial iterations. Mid-exposures' timestamps were transformed into barycentric Julian dates and barycentric dynamical time $\rm{BJD_{TDB}}$. Out-of-transit measurements with a trial transit model were used to try de-trending against air mass, time, the x and y position on the chip, and seeing. The final light curves were normalised to a baseline outside the transit.

\subsection{Ground-based out-of-transit monitoring}

\MakeTable{r l c c c c c c}{12.5cm}{Summary for the out-of-transit monitoring}
{\hline
\#  & Date UT & ${\rm BJD_{TDB}}$ start--end & $N_{\rm{obs}}$ & $t_{\rm{exp}}$ & $\Gamma$ & $pnr$  & $t_{\rm{run}}$ \\
    &         & (2450000+)                   &                & (s)            &          & (ppth) & (h)            \\
 \hline
\multicolumn{8}{c}{HAT-P-10}\\ 
 1 & 2018-10-13 & 8405.4893--8405.6636 & 503 & 27 & 2.00 & 1.30 & 4.18 \\
 2 & 2018-10-14 & 8406.4028--8406.6445 & 694 & 27 & 2.00 & 1.45 & 5.80 \\
   & \multicolumn{1}{c}{$\cdots$} & $\cdots$ & $\cdots$ & $\cdots$ & $\cdots$ & $\cdots$  & $\cdots$  \\
 9 & 2019-10-29 & 8786.2891--8786.6725 & 799 & 37 & 1.57 & 1.50 & 9.20 \\
\multicolumn{7}{r}{total:}                                 & 45.50\\ 
\multicolumn{8}{c}{HAT-P-17}\\ 
 1 & 2016-08-16 & 7625.3073--7625.5387 & 992 & 17 & 3.02 & 1.61 & 5.55 \\
 2 & 2016-08-17 & 7626.2990--7626.5497 & 1057 & 17 & 3.02 & 1.58 & 6.02 \\
   & \multicolumn{1}{c}{$\cdots$} & $\cdots$ & $\cdots$ & $\cdots$ & $\cdots$ & $\cdots$  & $\cdots$  \\
 7 & 2018-10-24 & 8416.2079--8416.4517 & 861 & 17 & 3.01 & 1.11 & 5.85 \\
\multicolumn{7}{r}{total:}                                 & 37.23\\ 
\multicolumn{8}{c}{HAT-P-44}\\ 
  1 & 2016-12-28 & 7751.4948--7751.6045 & 142 & 55 & 1.00 & 2.17 & 2.63 \\
  2 & 2017-01-02 & 7756.4872--7756.6363 & 201 & 55 & 1.00 & 2.58 & 3.58 \\
   & \multicolumn{1}{c}{$\cdots$} & $\cdots$ & $\cdots$ & $\cdots$ & $\cdots$ & $\cdots$  & $\cdots$  \\
 29 & 2019-04-07 & 8581.3000--8581.5916 & 618 & 37 & 1.50 & 2.36 & 7.00 \\
\multicolumn{7}{r}{total:}                                 & 167.30\\ 
\multicolumn{8}{c}{Qatar-6}\\ 
  1 & 2018-02-20 & 8170.4739--8170.5470 & 207 & 25 & 2.00 & 1.75 & 1.75 \\
  2 & 2018-03-17 & 8195.4027--8195.6521 & 709 & 25 & 2.00 & 1.96 & 5.99 \\
   & \multicolumn{1}{c}{$\cdots$} & $\cdots$ & $\cdots$ & $\cdots$ & $\cdots$ & $\cdots$  & $\cdots$  \\
 18 & 2019-05-05 & 8609.3391--8609.5531 & 457 & 37 & 1.50 & 1.56 & 5.14 \\
\multicolumn{7}{r}{total:}                                 & 96.90\\ 
\hline
\multicolumn{8}{l}{Designations as in Table 4. $t_{\rm{run}}$ is the time covered by uninterrupted observations.}\\ 
\multicolumn{8}{l}{This table is available in its entirety in a machine-readable form at the CDS. }  \\
\multicolumn{8}{l}{A portion is shown here for guidance regarding its form and content.}  \\
}

For four targets, HAT-P-10, HAT-P-17, HAT-P-44, and Qatar-6, additional photometric time series outside transits were acquired with the PIW 0.6 telescope. This photometric monitoring aimed to search for transit-like signals from hypothetical other planets in those systems before the TESS observations. More than 340 hours of observations were acquired in total between 2016 and 2019. The data reduction procedure was the same as for transit observations (Sect.~3.2). The details on individual runs are collected in Table~5.


\section{Data analysis and results}

\subsection{Transit modelling}

The transit parameters in the studied systems were redetermined using the new photometric data. For transits covered with TESS observations, the chunks of the length of 5 times the transit duration and centred at the expected mid-transit time were extracted from the SC time series. The LC data were skipped due to a lower time resolution, which dilutes transit shape (Hernandez Camero \etal 2023). The TESS light curves and the new ground-based observations were modelled simultaneously with the Transit Analysis Package (TAP, Gazak \etal 2012). Transit geometry was determined by the ratio of the planet to star radii $R_{\rm{p}}/R_{\star}$, semi-major axis scaled in star radii $a/R_{\star}$, and orbital inclination $i_{\rm{orb}}$. The mid-transit time $T_{\rm{mid}}$ was determined for each observed transit. Possible photometric trends in the time domain were accounted for with a second-order polynomial fitted for each light curve separately.

The pixel scale of the TESS cameras is 21 arc seconds per pixel, and photometric measurements may suffer from contaminating flux from nearby sources. Thus, we modified TAP by adding the flux contamination parameter $c_{\rm F}$, defined as  
\begin{equation}
	c_{\rm F} = \frac{\Delta F}{F_0} \, ,
\end{equation}
where $\Delta F$ is the additional flux in an aperture and $F_0$ is the unaffected target flux. This parameter was free for TESS light curves, and its value was common to all transits. For ground-based observations, $c_{\rm F}$ was fixed at zero.

The limb darkening (LD) was approximated with the quadratic form coded with $u_1$ and $u_2$ coefficients. As advocated by Patel \& Espinoza (2022), both parameters were free in the transit model with a separate pair for each passband. If a single light curve was acquired in a given filter, LD coefficients were allowed to vary around theoretical predictions from tables of Claret \& Bloemen (2011) under the Gaussian penalty with a conservative value of 0.1.

HAT-P-17~b is the only planet in our sample moving on a noncircular orbit. The values of its eccentricity $e_{\rm b} = 0.342 \pm 0.004$ and argument of periastron \mbox{$\omega_{\rm b} = 200.5^{\circ} \pm 1.3^{\circ}$} were taken from Bonomo \etal (2017) and allowed to vary under the Gaussian penalties equal to the uncertainties of those parameters. For remaining systems, the circular orbits were assumed following Bonomo \etal (2017) results or from the discovery papers.

The parameters of the best-fitting models and their uncertainties were derived with the Markov Chain Monte Carlo (MCMC) method, as described in detail in Maciejewski (2020). They are collected in Table~6. The LD coefficients are listed in Table~7. Figure~1 displays TESS's phase-folded transit light curves and the best-fitting models. The individual ground-based light curves are presented in Figs.~2 and 3.

\MakeTable{lccccc}{12.5cm}{System parameters from transit light curve modeling}
{\hline
Planet     & $R_{\rm{p}}/R_{\star}$       & $a/R_{\star}$             & $i_{\rm{orb}}$ $(^{\circ})$ & $c_{\rm F}$ $(\%)$ & $c_{\rm FluxCT}$ $(\%)$  \\
\hline
HAT-P-4 b  & $0.0874^{+0.0010}_{-0.0011}$ & $5.994^{+0.033}_{-0.063}$ & $89.09^{+0.63}_{-0.84}$ & $6.5^{+2.6}_{-2.8}$ & 0.6 \\
HAT-P-10 b & $0.1310^{+0.0019}_{-0.0017}$ & $12.12^{+0.16}_{-0.22}$   & $89.14^{+0.55}_{-0.47}$ & $5.1^{+2.2}_{-2.2}$ & 1.1 \\
HAT-P-12 b & $0.1418^{+0.0024}_{-0.0026}$ & $11.50^{+0.28}_{-0.26}$   & $88.45^{+0.57}_{-0.38}$ & $4.4^{+2.4}_{-2.6}$ & 0.1 \\
HAT-P-17 b & $0.1227^{+0.0021}_{-0.0021}$ & $22.71^{+0.37}_{-0.35}$   & $89.35^{+0.21}_{-0.16}$ & $6.5^{+3.4}_{-3.6}$ & 4.8 \\
HAT-P-19 b & $0.1389^{+0.0025}_{-0.0030}$ & $11.97^{+0.31}_{-0.25}$   & $88.31^{+0.53}_{-0.33}$ & $4.0^{+2.7}_{-2.9}$ & 0.1 \\
HAT-P-32 b & $0.1524^{+0.0009}_{-0.0009}$ & $6.051^{+0.051}_{-0.048}$ & $88.12^{+0.61}_{-0.42}$ & $2.3^{+1.0}_{-1.0}$ & 1.0 \\
HAT-P-44 b & $0.1371^{+0.0018}_{-0.0017}$ & $11.93^{+0.22}_{-0.24}$   & $88.86^{+0.62}_{-0.42}$ & $4.7^{+2.2}_{-2.3}$ & 0.8 \\
Qatar-6 b  & $0.1512^{+0.0049}_{-0.0033}$ & $12.30^{+0.15}_{-0.17}$   & $85.86^{+0.10}_{-0.12}$ & $-3.2^{+4.9}_{-5.5}$ & 1.0 \\
TrES-4 b   & $0.0976^{+0.0014}_{-0.0012}$ & $5.93^{+0.11}_{-0.11}$    & $82.54^{+0.25}_{-0.25}$ & $0.0^{\rm a}$ & 2.1 \\
WASP-48 b  & $0.0952^{+0.0012}_{-0.0013}$ & $4.569^{+0.081}_{-0.067}$ & $81.55^{+0.38}_{-0.31}$ & $2.7^{+2.9}_{-2.4}$ & 1.8 \\
\hline
\multicolumn{6}{l}{$c_{\rm FluxCT}$ is the total flux contamination predicted by FluxCT (Schonhut-Stasik \& Stassun 2023).}  \\
\multicolumn{6}{l}{$^{\rm a)}$ not fitted because only one ground-based light curve of moderate quality was available.}  \\
}

\MakeTable{lcccccc}{12.5cm}{Limb darkening coefficients determined from transit light curve modeling}
{\hline
System     & $u_{\rm 1,TESS}$  & $u_{\rm 2,TESS}$  & $u_{\rm 1,clear}$  & $u_{\rm 2,clear}$  & $u_{1,R}$  & $u_{2,R}$ \\
\hline
HAT-P-4  & $0.41^{+0.08}_{-0.08}$ & $0.07^{+0.14}_{-0.14}$ & $-$ & $-$ & $^{\rm a)}$$0.40^{+0.09}_{-0.09}$ & $^{\rm a)}$$0.17^{+0.14}_{-0.14}$  \\
HAT-P-10 & $0.48^{+0.09}_{-0.09}$ & $0.12^{+0.18}_{-0.18}$ & $-$ & $-$ & $^{\rm b)}$$0.56^{+0.10}_{-0.10}$ & $^{\rm b)}$$0.17^{+0.19}_{-0.18}$  \\
HAT-P-12 & $0.51^{+0.13}_{-0.13}$ & $0.01^{+0.27}_{-0.25}$ & $0.66^{+0.13}_{-0.13}$ & $-0.08^{+0.26}_{-0.23}$ & $-$ & $-$  \\
HAT-P-17 & $0.429^{+0.064}_{-0.061}$ & $0.19^{+0.13}_{-0.14}$ & $0.60^{+0.12}_{-0.13}$ & $0.11^{+0.21}_{-0.20}$ & $-$ & $-$  \\
HAT-P-19 & $0.37^{+0.19}_{-0.19}$ & $0.22^{+0.33}_{-0.34}$ & $0.50^{+0.15}_{-0.14}$ & $0.12^{+0.30}_{-0.30}$ & $-$ & $-$  \\
HAT-P-32 & $0.304^{+0.064}_{-0.064}$ & $0.06^{+0.12}_{-0.12}$ & $0.416^{+0.054}_{-0.054}$ & $0.00^{+0.10}_{-0.10}$ & $-$ & $-$ \\
HAT-P-44 & $0.50^{+0.13}_{-0.15}$ & $-0.10^{+0.27}_{-0.23}$ & $0.64^{+0.10}_{-0.10}$ & $-0.16^{+0.20}_{-0.19}$ & $-$ & $-$ \\
Qatar-6 & $0.50^{+0.32}_{-0.33}$ & $0.06^{+0.36}_{-0.32}$ & $0.38^{+0.22}_{-0.21}$ & $0.20^{+0.27}_{-0.31}$ & $-$ & $-$  \\
TrES-4  & $0.42^{+0.29}_{-0.27}$ & $0.00^{+0.31}_{-0.33}$ & $-$ & $-$ & $-$ & $-$ \\
WASP-48 & $0.23^{+0.20}_{-0.16}$ & $0.22^{+0.22}_{-0.26}$ & $0.43^{+0.28}_{-0.25}$ & $0.11^{+0.31}_{-0.36}$ & $-$ & $-$  \\
\hline
\multicolumn{7}{l}{$^{\rm a)}$ no distinction between $R_{\rm C}$ and Sloan-$r'$ because of their similar spectral bands.}  \\
\multicolumn{7}{l}{$^{\rm b)}$ to be specific: $R_{\rm C}$.}  \\
}

\begin{figure}[thb]
\begin{center}
\includegraphics[width=1.0\textwidth]{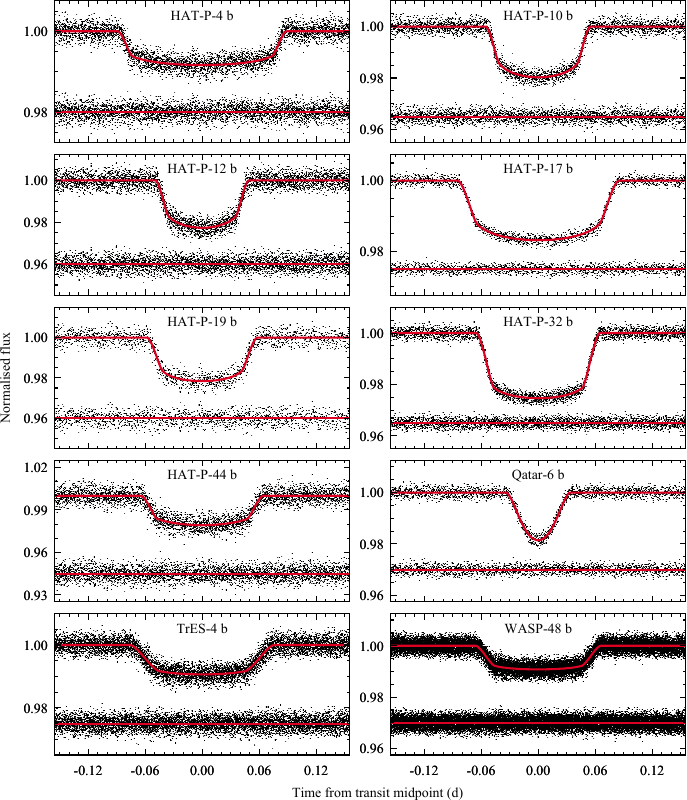}
\end{center}
\FigCap{Phase-folded transit light curves observed with TESS and the best-fitting models for the planets of our sample. The residuals are plotted below each light curve.}
\end{figure}

\begin{figure}[thb]
\begin{center}
\includegraphics[width=0.97\textwidth]{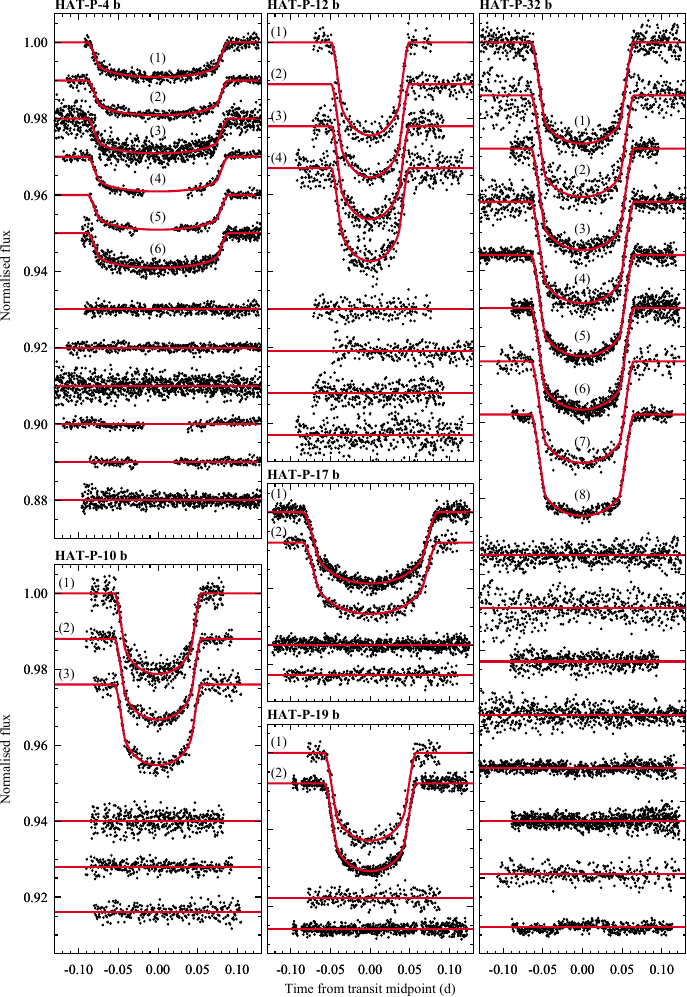}
\end{center}
\FigCap{New ground-based transit light curves for HAT-P-4~b, HAT-P-10~b, HAT-P-12~b, HAT-P-17~b, HAT-P-19~b, and HAT-P-32~b. Individual photometric time series are sorted by the observation date and labelled with ID numbers assigned in Table~4. The best-fitting models are drawn with red lines, and the residuals are plotted below.}
\end{figure}

\begin{figure}[thb]
\begin{center}
\includegraphics[width=1.0\textwidth]{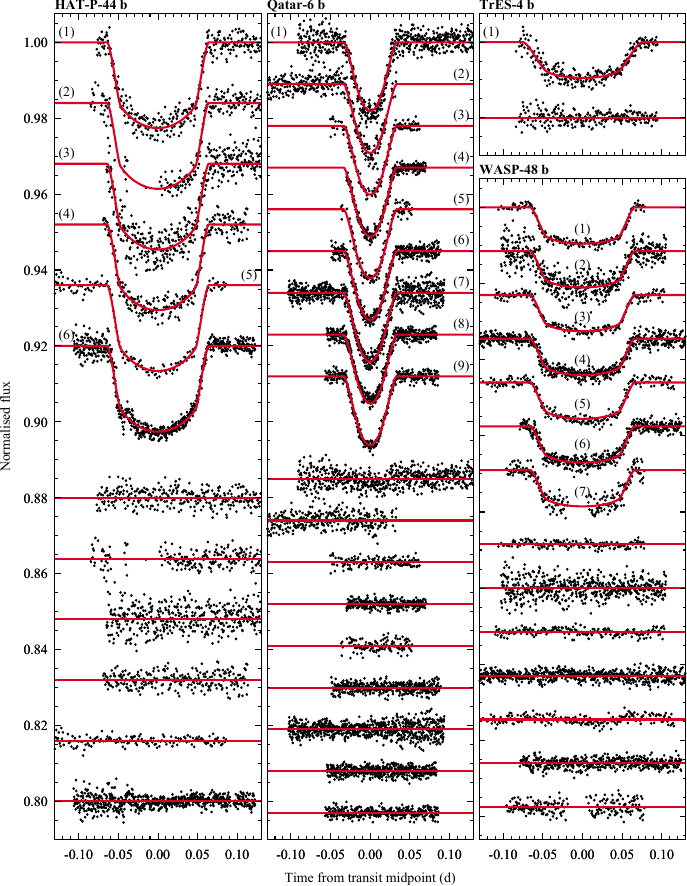}
\end{center}
\FigCap{Same as for Fig.~2, but for HAT-P-44~b, Qatar-6~b, TrES-4~b, and WASP-48~b.}
\end{figure}

For the homogeneity of our transit timing analysis, mid-transit times were also redetermined for the literature data. We only considered the light curves that are publicly available, repeating the fitting procedure with TAP. The new and redetermined mid-transit times are given in Table~8.

\MakeTable{ l c c c c l}{12.5cm}{Transit mid-points for the studied planets}
{\hline
Planet      & $E$ & $T_{\rm mid}$ (${\rm BJD_{TDB}}$) & $+\sigma$ (d) & $-\sigma$ (d) & Data source\\
\hline
HAT-P-4 b   &  $950$ & $2457149.512385$ & $0.000418$ & $0.000428$ & OSN 1.5\\
HAT-P-4 b   & $1057$ & $2457476.560467$ & $0.000462$ & $0.000495$ & OSN 1.5\\
HAT-P-4 b   & $1303$ & $2458228.464403$ & $0.000599$ & $0.000594$ & PIW 0.6\\
$\cdots$    & $\cdots$ & $\cdots$ & $\cdots$ & $\cdots$ & $\cdots$\\
WASP-48 b   & $2093$ & $2459851.184428$ & $0.001049$ & $0.001055$ & TESS\\
\hline
\multicolumn{6}{l}{This table is available in its entirety in a machine-readable form at the CDS. }  \\
\multicolumn{6}{l}{A portion is shown here for guidance regarding its form and content.}  \\
}

\subsection{Transit timing}

Transit timing data sets were used to refine transit ephemerides in the form:
\begin{equation}
     T_{\rm mid }(E) = T_0 + P_{\rm orb} \cdot E \, , \;
\end{equation}
where $E$ is the transit number counted from the reference epoch $T_0$, taken from the discovery papers. The best-fitting parameters and their $1\sigma$ uncertainties were extracted from posterior probability distributions produced by 100 MCMC walkers with $10^4$ steps each and the first 1000 steps discarded. The results are given in Table~9, and the transit timing residuals against the refined ephemerides are plotted in Figs.~4 and 5.

\MakeTable{ l c c c }{12.5cm}{Transit ephemeris elements for the investigated planets}
{\hline
Planet      & $T_0$ (${\rm BJD_{TDB}}$) & $P_{\rm orb}$ (d) & $\chi^2_{\rm red}$\\
\hline
HAT-P-4 b   & $2454245.81527 \pm 0.00041$ & $3.05652330 \pm 0.00000030$ & 1.01 \\
HAT-P-10 b  & $2454759.68719 \pm 0.00020$ & $3.72247955 \pm 0.00000020$ & 0.96 \\
HAT-P-12 b  & $2454216.77332 \pm 0.00016$ & $3.21305813 \pm 0.00000014$ & 1.68 \\
HAT-P-17 b  & $2454801.17059 \pm 0.00045$ & $10.3385344 \pm 0.0000010$  & 0.83 \\
HAT-P-19 b  & $2455091.53491 \pm 0.00035$ & $4.00878322 \pm 0.00000038$ & 0.90 \\
HAT-P-32 b  & $2454420.44699 \pm 0.00010$ & $2.150008111 \pm 0.000000056$ & 1.34 \\
HAT-P-44 b  & $2455696.93746 \pm 0.00036$ & $4.30119078 \pm 0.00000051$ & 0.88 \\
Qatar-6 b   & $2457784.03266 \pm 0.00016$ & $3.50620150 \pm 0.00000047$ & 0.97 \\
TrES-4 b    & $2454230.90558 \pm 0.00048$ & $3.55392915 \pm 0.00000037$ & 1.18 \\
WASP-48 b   & $2455364.55185 \pm 0.00023$ & $2.14363663 \pm 0.00000015$ & 1.01 \\
\hline
\multicolumn{4}{l}{$\chi^2_{\rm red}$ is the reduced chi-square for the refined linear ephemeris}  \\
}

\begin{figure}[thb]
\begin{center}
\includegraphics[width=1.0\textwidth]{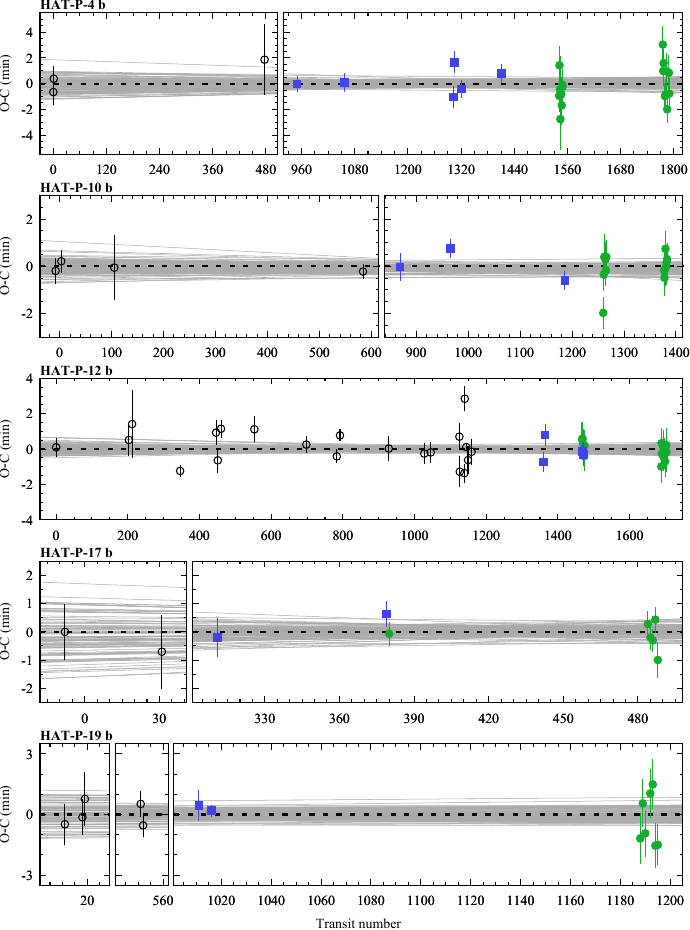}
\end{center}
\FigCap{Transit-timing residuals against the refined linear ephemerides for HAT-P-4~b, HAT-P-10~b, HAT-P-12~b, HAT-P-17~b, and HAT-P-19~b. The values from the TESS photometry and ground-based observations are marked with green dots and blue squares, respectively. The redetermined literature values are plotted with open circles. Dashed lines mark the zero value. The ephemeris uncertainties are illustrated by grey lines plotted for 100 sets of parameters drawn from the Markov chains.}
\end{figure}

\begin{figure}[thb]
\begin{center}
\includegraphics[width=1.0\textwidth]{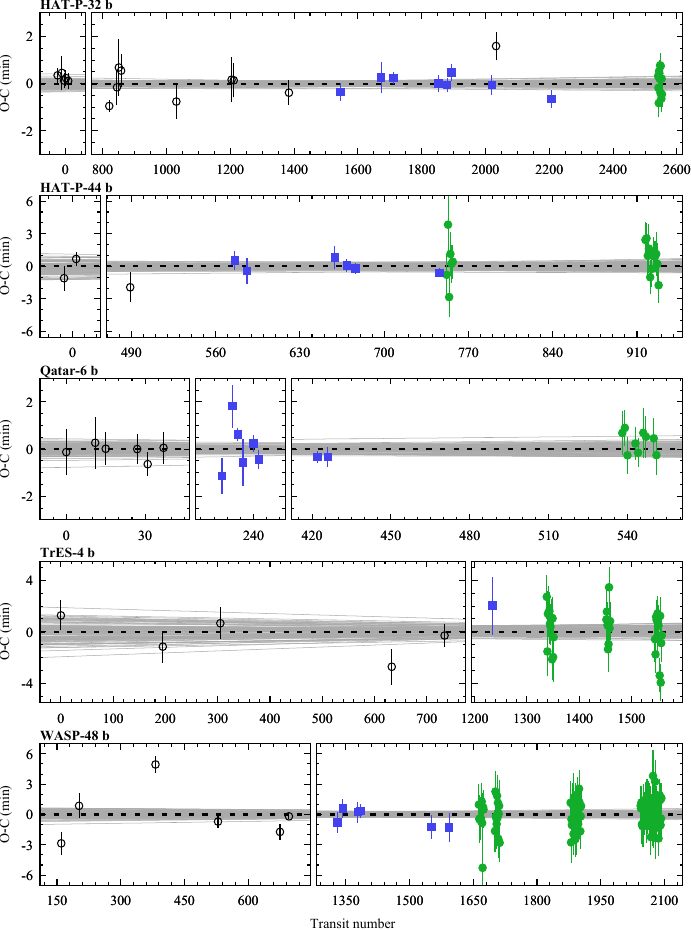}
\end{center}
\FigCap{The same as Fig.~4 but for HAT-P-32~b, HAT-P-44~b, Qatar-6~b, TrES-4~b, and WASP-48~b.}
\end{figure}

The mid-transit times were probed for possible long-term trends which could be caused by a monotonic or periodic change of $P_{\rm orb}$. Trail quadratic ephemerides in the form:
\begin{equation}
 T_{\rm{mid}}= T_0 + P_{\rm{orb}} \cdot E + \frac{1}{2} \frac{{\rm d} P_{\rm{orb}}}{{\rm d} E} \cdot E^2 \, , \;
\end{equation}
where ${{\rm d} P_{\rm{orb}}}/{{\rm d} E}$ is the change in the orbital period between succeeding transits, were evaluated. The Bayesian information criterion (BIC) disfavours the quadratic ephemerides for all planets of our sample. For four systems, HAT-P-4, HAT-P-10, HAT-P-19, and HAT-P-32, the radial acceleration of the barycentre, $\dot{\gamma}$, was detected in the Doppler measurements (Bonomo \etal 2017). For those systems, Table~10 lists the derived values of ${{\rm d} P_{\rm{orb}}}/{{\rm d} E}$ and the predicted values of $\left( {{\rm d} P_{\rm{orb}}}/{{\rm d} E} \right)_{\rm RV}$, which were calculated with the formula:  
\begin{equation}
 \left( \frac{{\rm d} P_{\rm{orb}}}{{\rm d} E} \right)_{\rm RV} = \frac{\dot{\gamma}}{c} P_{\rm{orb}}^2 \, . \;
\end{equation}
The current timing data sets do not verify the RV predictions for HAT-P-4 and \mbox{HAT-P-10}. They were consistent within the 1$\sigma$ level due to the high relative errors of ${{\rm d} P_{\rm{orb}}}/{{\rm d} E}$. Interestingly, the discrepant results at 6.3$\sigma$ and 4.4$\sigma$ were obtained for HAT-P-19 and HAT-P-32, respectively.   

\MakeTable{ l c c }{12.5cm}{Constraints on a constant period change from transit timing and radial acceleration}
{\hline
System      & $\frac{{\rm d} P_{\rm{orb}}}{{\rm d} E}$ $(10^{-10}\,{\rm d})$ & $\left( \frac{{\rm d} P_{\rm{orb}}}{{\rm d} E} \right)_{\rm RV}$ $(10^{-10}\,{\rm d})$\\
\hline
HAT-P-4   & $-1.5 \pm 9.5$ & $7.0 \pm 1.0$ \\
HAT-P-10  & $-0.4 \pm 9.8$ & $-9.2 \pm 1.3$ \\
HAT-P-19  & $-13 \pm 23$ & $240 \pm 33$ \\
HAT-P-32  & $2.6 \pm 1.5$ & $-14.5 \pm 3.5$ \\
\hline
}

The linear ephemerides were subtracted from the timing data, and the residuals were searched for periodic signals employing the analysis of variance algorithm (AoV, Schwarzenberg-Czerny 1996). Periodograms were calculated for trial periods between $2$ and $10^4$ transit intervals for each planet. The empirical levels of the false alarm probability (FAP) were determined with the bootstrap method, which was based on $10^5$ trials. As shown in Fig.~6, no statistically significant signal was detected for any planet.

\begin{figure}[thb]
\begin{center}
\includegraphics[width=1.0\textwidth]{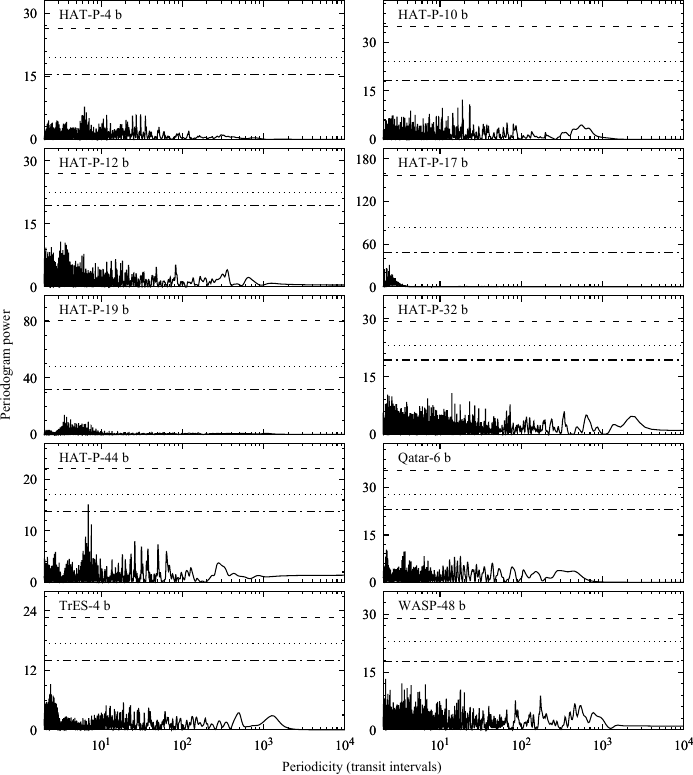}
\end{center}
\FigCap{AoV periodograms for the timing residuals against the redetermined ephemerides. The dashed and dotted horizontal lines show the empirical FAP levels of 5\%, 1\%, and 0.1\% (from the bottom up).}
\end{figure}

\subsection{Search for additional transiting planets}

\begin{figure}[thb]
\begin{center}
\includegraphics[width=0.97\textwidth]{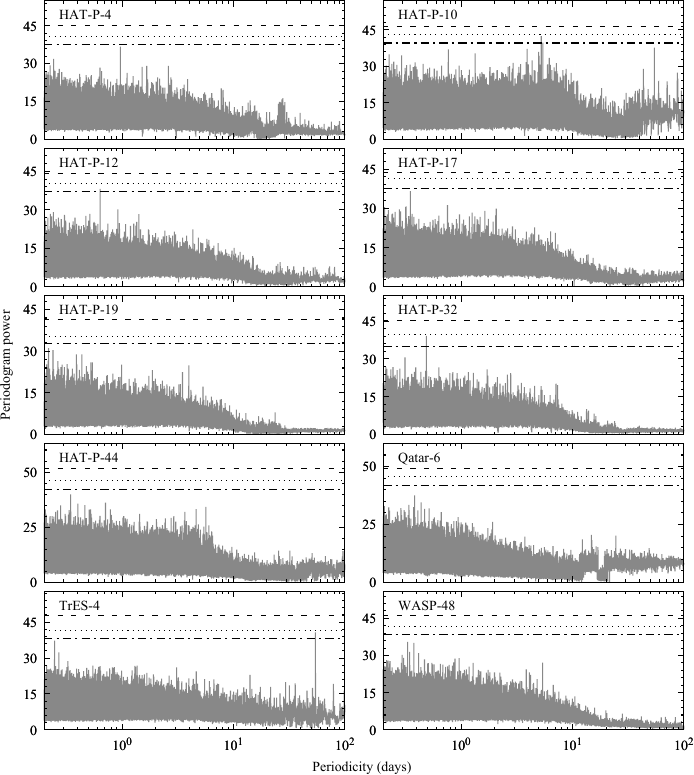}
\end{center}
\FigCap{AoVtr periodograms for out-of-transit observations in the examined systems. The dashed and dotted horizontal lines mark the empirical FAP levels of 5\%, 1\%, and 0.1\% (from the bottom up).}
\end{figure}

Adopting the procedure from Maciejewski (2020), we used the AoVtr code (Schwarzenberg-Czerny \& Beaulieu 2006) to search for transit-like flux drops in the joint SC and LC TESS photometric time series. For HAT-P-10, HAT-P-17, HAT-P-44, and Qatar-6, the light curves were enhanced with ground-based photometric monitoring (Sect.~3.3). The refined transit ephemerides were used to mask out transits and occultations of the known giant planets. The algorithm was run for trial periods between $0.2$ and $100$ days with a resolution in the frequency domain equal to $5 \times 10^{-5}$ day$^{-1}$. As the algorithm's sensitivity might depend on the number of bins of a phase-folded light curve, the procedure was iterated over the number of bins from 10 to 100 with a step of 10. The periodogram with the highest peak was saved for further analysis. The statistical significance of peaks was estimated with the bootstrap method executed on $10^4$ resampled datasets. The periodograms with the FAP levels are plotted in Fig.~7.

No statistically significant signal, \ie with FAP below 0.1\%, was detected for any system. For three systems, \ie HAT-P-10, HAT-P-32, and TrES-4, the strongest peaks reached the FAP level close to 1\%. Thus, we visually inspected the phase-folded light curves to verify these signals. We finally identified them as not actual transit shapes.

Injection-recovery tests were adopted from Maciejewski (2020) to determine the transit detection thresholds for the individual systems. The transit depths were then converted into the upper radii of potential planets that remain below the detection thresholds of the present photometric time series. The results are displayed in Fig.~8.

\begin{figure}[thb]
\begin{center}
\includegraphics[width=0.97\textwidth]{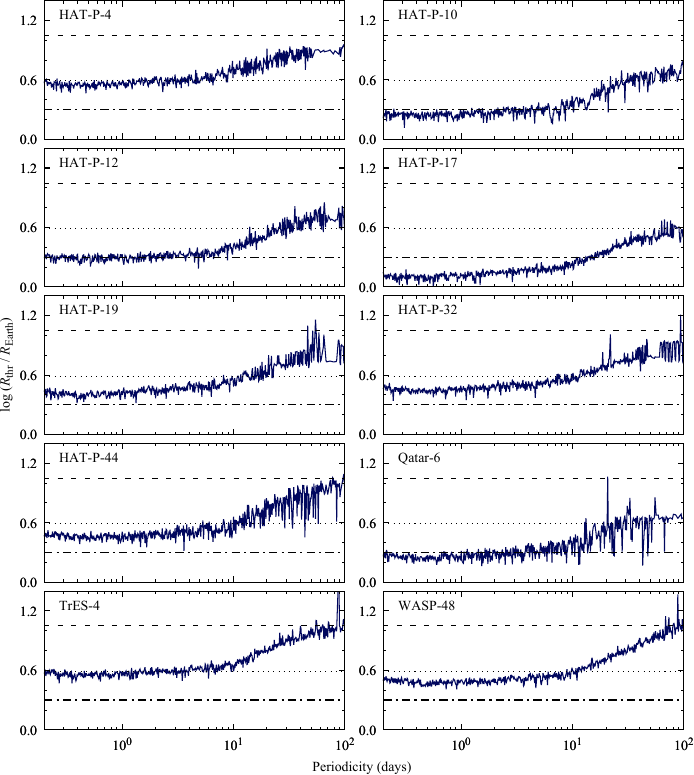}
\end{center}
\FigCap{Empirical upper constraints on radii of hypothetical planets that remain below the transit detection limit in the investigated systems. The dashed and dotted horizontal lines mark the values for 2 $R_{\oplus}$, Neptune, and Jupiter (from the bottom up).}
\end{figure}

\subsection{Variable stars in the HAT-P-17 field}

Whilst reducing the ground-based observations, we identified four multi-periodic pulsating variable stars in the field around HAT-P-17. We extracted their light curves from TESS data for further analysis. We list those variables in Table~11 and the detected pulsation frequencies in Table~12. Their exemplary light curves are plotted in Fig.~9. Below, we give their brief characteristics.

\MakeTable{ c l l l c }{12.5cm}{Short-period pulsating variable stars identified in the field of HAT-P-17.}
{\hline
ID  & Name                          & RA (J2000)  & Dec (J2000)   & $m_{\rm G}$ \\
    &                               & hh:mm:ss.s  & $\pm$dd:mm:ss & (mag)        \\
\hline
V1  & BD+30 4487                    & 21:36:50.1  & +30:41:01.4   & 10.68  \\
V2  & Gaia DR3 1849720750852486656  & 21:39:17.7  & +30:16:12.1   & 12.95  \\
V3  & TYC 2717-453-1                & 21:38:22.2  & +30:33:22.3   & 11.89  \\
V4  & Gaia DR3 1849743737517511296  & 21:39:06.9  & +30:28:18.6   & 13.34  \\
\hline
\multicolumn{5}{l}{Coordinates and apparent brightness in the Gaia $G$ band $m_{\rm G}$ are taken from the DR3.}  \\
}

\textit{V1}. The star was found to display a 0.08-d periodic variation with a range of 2\%. It was observed with TESS in Sector 55 with a cadence of 10 minutes. The periodogram analysis performed with the PERIOD04 package (Lenz and Breger 2005) revealed a wealth of periodicities, demonstrating that this is a multi-periodic $\delta$ Scuti star. Using a standard pre-whitening procedure, we identified 14 frequencies with a signal-to-noise ratio (S/N) above 5.4, as Baran \etal (2015) advocate for time series data from space missions. The star has not been previously reported to be variable. 

\textit{V2}. A rapid brightness modulation with a period of 0.037 d and a range floating up to 2\% was observed. The periodogram analysis of the TESS data, acquired in sector 56 with the 200-s cadence, revealed 8 frequencies between 22.9 and \mbox{28.3 day$^{-1}$} with amplitudes up to 0.3\%. The star is a multi-periodic $\delta$ Scuti variable. No reports on its variability can be found in the literature. 

\textit{V3}. We initially identified this star as a single-period 0.026-d pulsator changing its brightness by 2\%. It is registered in the International Variable Star Index (VSX) and assigned a $\delta$ Scuti type. The TESS observations, obtained in Sector 56 with the 200 s cadence, allowed us to identify other 3 frequencies with much lower amplitudes. Therefore, we classify this star as a multi-periodic $\delta$ Scuti star.

\textit{V4}. Brightness variations with a period of 0.093 d and a peak-to-peak amplitude of 4\% were found for this star. It is classified in VSX as a $\delta$ Scuti variable and is registered in the Czech Variable Star Catalogue as CzeV1766. Our ground-based observations provided a hint for an amplitude modulation which was confirmed in the 200-s cadence TESS light curve from Sector 56. The power spectrum shows 16 frequencies between 10.1 and 28.2 day$^{-1}$ and amplitudes up to 1\%.
 
\begin{figure}[thb]
\begin{center}
\includegraphics[width=0.97\textwidth]{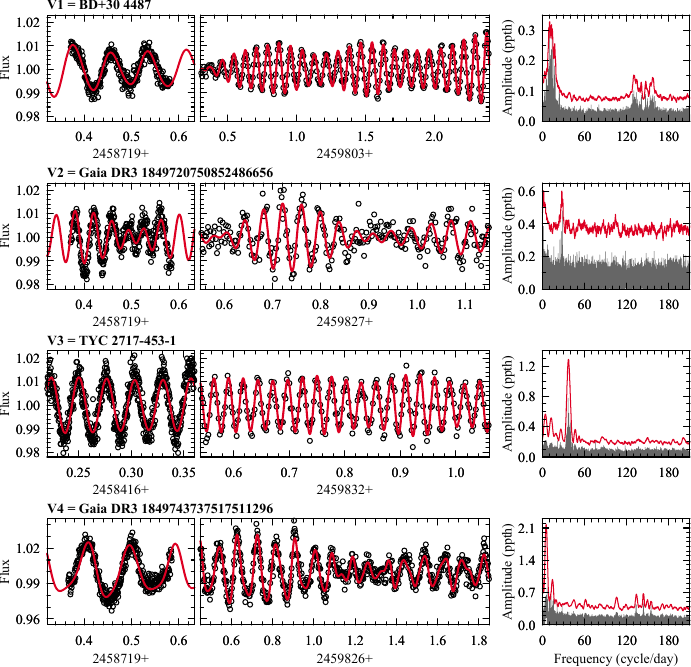}
\end{center}
\FigCap{Exemplary light curves for the pulsating variable stars observed in the field of HAT-P-17. Ground-based light curves observed by us are displayed in the left column. Portions of TESS observations are plotted in the middle column. They were selected to probe the course of variability best. Red lines show the best-fitting models, which were obtained for joint data sets. The periodograms of the final residuals are presented in the right column. The detection threshold of conservative S/N = 5.4 is plotted with red lines.}
\end{figure}

\MakeTable{ l c c c c }{12.5cm}{Multi-frequency solutions for the pulsating stars.}
{\hline
     & V1                     & V2                     & V3                     & V4                     \\
No.  & Frequency (day$^{-1}$) & Frequency (day$^{-1}$) & Frequency (day$^{-1}$) & Frequency (day$^{-1}$) \\
     & Amplitude (ppth)       & Amplitude (ppth)       & Amplitude (ppth)       & Amplitude (ppth)       \\
\hline
F1  & $11.99059(5)$  & $26.81479(34)$ & $37.6674986(12)$ & $10.790088(22)$ \\
    & $7.250(14)$    & $3.351(57)$    & $11.153(34)$     & $9.85(8)$ \\ 
F2  & $11.82779(7)$  & $25.60031(37)$ & $43.735803(14)$  & $11.245132(17)$ \\
    & $3.743(16)$    & $3.002(57)$    & $1.257(33)$      & $8.630(64)$  \\
F3  & $13.18320(47)$ & $22.9902(38)$  & $42.816332(46)$  & $10.022355(16)$ \\
    & $2.260(16)$    & $2.36(94)$     & $0.447(33)$      & $7.78(7)$ \\
F4  & $11.36592(11)$ & $28.25227(52)$ & $42.892795(38)$  & $17.40692(16)$ \\
    & $2.149(18)$    & $2.167(56)$    & $0.529(33)$      & $3.305(65)$  \\
F5  & $ 9.49050(30)$ & $24.87537(86)$ &                  & $17.34429(31)$ \\
    & $1.113(17)$    & $1.43(6)$      &                  & $1.62(7)$  \\
F6  & $ 9.35757(28)$ & $22.9469(23)$  &                  & $10.81111(45)$ \\
    & $1.009(18)$    & $1.38(33)$     &                  & $1.57(8)$ \\
F7  & $11.663(9)  $  & $24.8129(16)$  &                  & $21.58108(5)$ \\
    & $0.72(12) $    & $0.75(6)$      &                  & $1.281(72)$  \\
F8  & $12.600(1)  $  & $23.025(37)$   &                  & $16.8804(28)$ \\
    & $0.574(18)$    & $0.6(1)$       &                  & $1.23(10)$ \\
F9  & $10.0924(6) $  &                &                  & $10.18861(61)$ \\
    & $0.502(17)$    &                &                  & $1.08(7)$  \\
F10 & $ 9.9923(11) $ &                &                  & $16.0711(2)$ \\
    & $0.423(19)$    &                &                  & $1.082(63)$ \\
F11 & $23.8183(9) $  &                &                  & $21.26679(44)$ \\
    & $0.237(15)$    &                &                  & $0.95(7)$  \\
F12 & $23.9803(41) $ &                &                  & $10.110(16)$ \\
    & $0.179(27)$    &                &                  & $0.92(14)$ \\
F13 & $ 5.32494(61)$ &                &                  & $10.703(37)$ \\
    & $0.192(18)$    &                &                  & $0.80(15)$ \\
F14 & $25.17385(80)$ &                &                  & $28.198905(44)$ \\
    & $0.146(15)$    &                &                  & $0.735(62)$ \\
F15 &                &                &                  & $21.6010(8)$ \\
    &                &                &                  & $0.81(8)$ \\
F16 &                &                &                  & $22.04(5)$ \\
    &                &                &                  & $0.74(13)$ \\
 \hline
\multicolumn{5}{l}{The best-fitting parameters come from the least-squares calculations. Their uncertainties}  \\
\multicolumn{5}{l}{are given in parentheses, estimated using Monte Carlo simulations based on 100 processes}  \\
\multicolumn{5}{l}{with $10^4$ steps each.}  \\
}


\section{Discussion}

The transit parameters re-determined by us are mostly in 1-2$\sigma$ agreement with previous works. For HAT-P-44, we provide the first verification of the transit parameters. For HAT-P-4~b, our value of $R_{\rm{p}}/R_{\star}$ agrees with the determinations of Christiansen \etal (2011) and Winn \etal (2011) within 0.2 and 0.4$\sigma$, respectively, while it is greater than the values of Kov\'acs \etal (2007) and Wang \etal (2021) by 4.7 and 3.1$\sigma$, respectively. In those two studies, transits of HAT-P-4~b appear to be shallower, implying the flux contamination of $\approx$12\%. Even in the wide TESS aperture, the expected contamination is much lower, \ie $\approx$0.6\%. In addition, our tests show that keeping the LD coefficients free does not overestimate $R_{\rm{p}}/R_{\star}$. For TrES-4~b, our value of $R_{\rm{p}}/R_{\star}$ agrees with the determinations by Mandushev \etal (2007), Sozzetti \etal (2009), and Chan \etal (2011) within 1$\sigma$. However, Sozzetti \etal (2015) reported deeper transits and hence the greater radius for the planet, deviating from our result by 4.1$\sigma$. To investigate the source of this discrepancy, we re-analysed the original light curves of Sozzetti \etal (2015). The apparently deeper transits result from the simplifications adopted by those authors: usage of a constant flux baseline outside the transits in conjunction with a relatively short coverage of out-of-transit observations and a simplified LD law in the linear form. Our approach applied to those light curves yields a value of $R_{\rm{p}}/R_{\star}$, which is consistent with other studies.

Our transit timing analysis revealed no sign of deviation from the linear ephemeris for each system. For two systems, HAT-P-19 and HAT-P-32, the RV accelerations claimed in the literature would produce apparent shortening or lengthening of the orbital period due to the light travel time (LTT) effect (Irwin 1952). Wide stellar companions in these systems can be excluded due to multiplicity studies of exoplanet host stars (see \eg Mugrauer 2019, Michel and Mugrauer 2021). For HAT-P-19, the outward acceleration of the systemic barycentre \mbox{$\dot{\gamma} = 0.440\pm 0.061 \; {\rm m \, s^{-1} \, day^{-1}}$} was detected by Hartman \etal (2011a) and confirmed in the homogenous RV analysis by Bonomo \etal (2017). The phenomenon of this magnitude would manifest as an apparent lengthening of the orbital period, giving a cumulative departure from the linear transit ephemeris by $\approx24$ minutes over 13 years, \ie the coverage of transit timing observations. We can discard the constant accelerations with $\dot{\gamma} > 0.058 \; {\rm m \, s^{-1} \, day^{-1}}$ and $\dot{\gamma} < -0.108 \; {\rm m \, s^{-1} \, day^{-1}}$ at 95\% confidence. The reported RV slope must be a fragment of a periodic signal shorter than the span of transit timing observations. Our reanalysis of RV measurements from Hartman \etal (2011a) reveals a planetary signal with an amplitude of $\approx 30 \; {\rm m \, s^{-1}}$, translating into the mass of the planetary companion of $\approx$$0.9 \: M_{\rm Jup} / \sin i$. The RV data set, however, places weak constraints on the signal's period, which is correlated with an orbital eccentricity: from 260 days for a circular orbit to tens of thousands of days for orbital eccentricities above 0.9.

We performed an analogous analysis for HAT-P-32, for which $\dot{\gamma} = -0.094\pm 0.023 \; {\rm m \, s^{-1} \, day^{-1}}$ was reported (Knutson \etal 2014, Bonomo \etal 2017). Thanks to the span of 13 years, the HAT-P-32~b's transit observations allowed us to discards constant accelerations with $\dot{\gamma} < -0.0007 \; {\rm m \, s^{-1} \, day^{-1}}$ and $\dot{\gamma} > 0.010 \; {\rm m \, s^{-1} \, day^{-1}}$ at 95\% confidence. The only RV data set was acquired by Knutson \etal (2014) between 2008 and 2012, providing a time coverage of about 1800 days. Thus, the orbital period of a third body in the system remains poorly constrained. Our numerical experiments show that a low-mass brown dwarf companion ($M_{\rm c} \sin i_{\rm c} \approx 20 \; M_{\rm Jup}$) would produce a detectable TTV signal with a period of about 17 years and a peak-to-peak amplitude of 2 minutes. Thus, the companion is rather a massive planet on a 4--6 au orbit.

The wide orbit companions to hot Jupiters in such planetary systems as \mbox{HAT-P-19} and HAT-P-32 are potentially responsible for the migration mechanism. Knowing their nature would be meaningful for theories of the formation of systems with hot Jupiters.

For the remaining systems with non-zero Doppler acceleration, \ie HAT-P-4 and HAT-P-10, the constraints coming from transit timing are too weak to address the values of $\dot{\gamma}$ derived from the RV measurements. As summarised in Table~10, our values of ${{\rm d} P_{\rm{orb}}}/{{\rm d} E}$ are consistent within 1$\sigma$ with the RV accelerations, as well as with zero.

For all of the planets of our sample, the flux contamination in the TESS aperture was found to be statistically indistinguishable from 0. This finding aligns with the values extracted with the online tool FluxCL (Schonhut-Stasik \& Stassun 2023), which are listed in Table~6. For the majority of the systems, these contaminations are at the level of 2\% or lower, so they are negligible. The only exception is \mbox{HAT-P-17}, for which $c_{\rm FluxCT} = 4.8\%$. Our determination of $c_{\rm F} = 6.5^{+3.4}_{-3.6}\%$ agrees with that value within 1$\sigma$.

Our analysis reveals no transiting planetary companions to the hot Jupiters of our sample. These planets were also found to be perfect clocks, with their transits following the linear ephemerides. This lack of resonant planetary companions completes the picture of the loneliness of hot Jupiters. However, Sariya \etal (2021) postulated the presence of a possible non-sinusoidal TTV for HAT-P-12~b, contrary to the conclusions of \"{O}zt\"{u}rk \& Erdem (2019). These perturbations would be induced by a 0.2 $M_{\rm Jup}$ companion on an 8.8-day orbit. Whilst this model was found to improve the value of the reduced $\chi^2$ for the transit timing data set of Sariya \etal (2021), we note that it devastates a single-planet RV solution. The purported companion would produce an RV signal with an amplitude similar to that of \mbox{HAT-P-12~b} that is not supported by the RV data. Our best-fitting linear ephemeris has $\chi^2_{\rm red} = 1.7$ and is overestimated by two outlying points from the literature, one from Alexoudi \etal (2018) and another from Mancini \etal (2018). Rejecting them causes the $\chi^2_{\rm red}$ value to drop to 1.0. Our investigation failed to identify the reason why those points stand out.

The negative results of our search for nearby planetary companions to hot Jupiters add to the non-detections already discussed in the literature (e.g., Hord \etal 2021, Wang \etal 2021). The known companions to hot Jupiters, including a most recently discovered super-Earth, WASP-132 c (Hord \etal 2022), are smaller than Neptune. The sensitivity of transit detection is related to, among other factors, the photometric precision, the amount of photometric data, and the size of the host star. We went down to Neptune sizes for HAT-P-4 and TrES-4 and entered the super-Earth regime in the HAT-P-17 system. For the remaining systems, we probed down to mini-Neptunes.

\section{Conclusions}

As the transit timing data discard the constant acceleration scenario for HAT-P-19~b and HAT-P-32~b, their systems may contain additional planets on wide orbits. Precise Doppler follow-up studies could confirm their existence. Transit times for HAT-P-12~b were consistent with the constant period model, discarding the non-sinusoidal TTV signal, which was recently claimed in the literature. The loneliness of the hot Jupiters of our sample supports the high-eccentricity migration as a primary path leading to the formation of systems with massive planets stripped of any close-in planetary companions.


\Acknow{We would like to thank all participants involved in the observations, especially F.~Hildebrandt. GM acknowledges the financial support from the National Science Centre, Poland through grant no. 2016/23/B/ST9/00579. MF and PJA acknowledge financial support from grants PID2019-109522GB-C52/AEI/ 10.13039/501100011033 of the Spanish Ministry of Science and Innovation (MICINN) and PY20\_00737 from Junta de Andaluc\'{\i}a. MF, AS, and PJA acknowledge financial support from the grant CEX2021-001131-S funded by MCIN/AEI/10.13039/ 501100011033. RB and MM acknowledge the support of the DFG priority program SPP 1992 ``Exploring the Diversity of Extrasolar Planets'' in projects NE515/58-1 and MU2695/27-1. This paper includes data collected with the TESS mission, obtained from the MAST data archive at the Space Telescope Science Institute (STScI). Funding for the TESS mission is provided by the NASA Explorer Program. STScI is operated by the Association of Universities for Research in Astronomy, Inc., under NASA contract NAS 5-26555. This research made use of Lightkurve, a Python package for Kepler and TESS data analysis (Lightkurve Collaboration, 2018). This research has made use of the SIMBAD database and the VizieR catalogue access tool, operated at CDS, Strasbourg, France, and NASA's Astrophysics Data System Bibliographic Services. This research has made use of the International Variable Star Index (VSX) database, operated at AAVSO, Cambridge, Massachusetts, USA. This project has received funding from the European Union's Horizon 2020 research and innovation programme under grant agreement No 730890. This material reflects only the authors views and the Commission is not liable for any use that may be made of the information contained therein.}


\end{document}